\documentclass[a4paper,10pt]{article}
\usepackage{graphicx}

\begin{document}


\title{Alternative Astronomical FITS imaging\footnote{Published in The 11th Asian-Pacific Regional IAU Meeting 2011, NARIT Conference Series, Vol. 1, S. Komonjinda, Y. Kovalev, and D. Ruffolo, eds.}}

\author{Eleni E. Varsaki$^1$, Nectaria A. B. Gizani$^2$\footnote{Email: ngizani@eap.gr}, Vassilis Fotopoulos$^1$\\ and Athanassios N. Skodras$^1$\\
$^1$ Digital Systems \& Media Computing Laboratory, School of Science and
Technology,\\ Hellenic Open University, Patra, Greece  \\
\noindent
$^2$ Physics Laboratory, School of Science and Technology, Hellenic Open University, \\ Patra, Greece \\}


\date{}
\maketitle

\paragraph{Abstract}

Astronomical radio maps are presented mainly in FITS format.
Astronomical Image Processing Software (AIPS) uses a set of tables attached
to the output map to include all sorts of information concerning the
production of the image. However this information together with information
on the flux and noise of the map is lost as soon as the image of the radio source in fits or other format is extracted from AIPS. This information would have been valuable to another astronomer who just uses NED, for example, to
download the map. In the current work, we show a method of data hiding
inside the radio map, which can be preserved under transformations, even for
example while the format of the map is changed from fits to other lossless
available image formats. \\

\section{Introduction}

Information manipulation to FITS (Flexible Image Transport System) file format
is applied by AIPS, using keywords able to handle original data and history
tracks. However certain measurements made by radio astronomers are not
saved into the file itself, like root mean square (rms) or flux. To get such
important information astronomers need to repeat the analysis of the data in
most cases. Considering this difficulty, we propose a data embedding algorithm
in order to save additional information into the FITS map. Data hiding is a well
known computer research field, with the purpose to embed information [1] into
a digital file, not into the header, but in a way that no human observer notices
the existence of such additional information. Data hiding finds applications
in authentication, tamper-proofing, copyright protection, secret communication
and other [2].

In this study we propose a new application of steganography, since additional
information can be embedded into the image itself, enabling the following user
to retain it after a file format conversion. This information can be extracted
from the FITS map and can be also received from the image pixels when FITS is
converted into TIF image format. Such application that converts FITS to TIFF
image format is ESA/ESO/NASA FITS Liberator 3 [3], a standalone application
released on October 2010.

\section{Procedure}

The proposed scheme consists of the embedding, format conversion and the
extractor. The input is the original FITS image and the message, that is
the embedded information. The output is the extracted message, that is the
information we extract from the TIF image format. The proposed data hiding
algorithm is based on the discrete Walsh-Hadamard Transform (WHT), where
selected WHT coefficients are changed according to the embedded bits. The
WHT is a non-sinusoidal, orthogonal transformation that decomposes a signal
into a set of orthogonal, rectangular waveforms called Walsh functions [4].
During embedding the original image is divided into 4$\times$4 non-overlapping blocks and WHT is applied to every block of image. The right-down WHT coefficient is read together with the message bit. One message bit is embedded into every block of image. $'1'$ requires positive WHT coefficient, $'0'$ requires negative WHT coefficient. If this is not the case then the coefficient's sign is changed and its value is shifted by a constant value d, which is experimentally chosen. Stego-FITS produced image is saved and converted into 16-bit TIF format, using ESA/ESO/NASA FITS Liberator 3. Then stego-TIF image enters the extractor, which reads the WHT coefficients of the stego-TIFF and extracts the message bits equal to $'1'$, if the coefficient is positive and $'0'$ otherwise.

\section{Results and Discussion}

Experiments have been implemented in Matlab 2010 release using the 16-bit
double precision floating point real numbers radio map CYGUS A. The message
size is 42976 bits, equal to the image's capacity, that is maximum amount of
data that can be embedded into the image map. The constant value d, according
to which the WHD coefficient is shifted is experimentally chosen to be 0:0001.
The proposed embedding algorithm is proved to be robust to FITS Liberator
conversion into 16-bit depth, by only using the linear stretch function with white and black levels equal to the maximum and minimum values of the produced
stego-fits image. As a consequence, all the dynamic range of the FITS image is
exported into the TIF image and further image editing can be made with any
image editing application, supporting 16-bit TIF format, with the advantage that
information can be extracted from the image that in any other case would have
been lost (eg. the flux of the source and the rms of the map). Experimental
results show that embedding procedure makes no visual artifacts, so that the
statistics remain unchanged and the good quality of the image is retained.

\section{References}
\noindent
[1] Husrev Sencar, Mahalingam Ramkumar, Ali Akansu, $'$Data Hiding Fundamentals and Applications$'$, Elsevier, 2004 \\
\noindent
[2] W. Bender, W. Butera, D. Gruhl, R. Hwang, F. J. Paiz, S. Pogreb, $'$Applications for data hiding$'$, IBM Systems Journal, Volume 39 Issue 3-4, July 2000 \\
\noindent
[3] The ESA/ESO/NASA FITS Liberator v3 - ESA/Hubble {\tt http://www.spacetelescope.org/projects/fits$_{-}$liberator/} \\
\noindent
[4] George J. Miao, Mark A. Clements, $'$Digital Signal Processing and Statistical Classification$'$, Artech House, Inc. Norwood, MA, USA, 2002 \\

\end{document}